# Impact of Chirality on the Properties of Two-Dimensional Images Propagating Through a Chiral Dispersive Thick Lens

**Salaheddeen Bugoffa**
Department of Engineering Technology
Norfolk State University
700 park Ave, Norfolk, VA, 23504
Emai: sgbugoffa@nsu.edu

**Hussin Ragb**
Electrical and Computer Engineering Department
Christian Brothers University
650 E Pkwy S, Memphis, TN, 38104
Email: hragb@cbu.edu

**ABSTRACT**

Dual image formation for a two-dimensional object via bimodal propagation through chiral-dispersive thick lens is derived. In this article, first-order frequency-dependent material dispersion of the dielectric permittivity and the lens material being chiral are considered. In addition, the thick lens is configured in a uniform background. A salient feature of a chiral thick lens is the inherent bimodal propagation via circular polarizations. Under chirality, two sets of ABCD frequency dependent matrices are derived for right- and left-circularly polarized modes based on standard paraxial and meridional conditions. For imaging purposes, a simple2D colored transparency is placed as an object before the thick lens. The image transmission across the lens examined via the ABCD matrix parameters and defocusing effects due to dispersion under different chirality bands.

## 1. Introduction

In a chiral medium, the propagating plane of electromagnetic (EM) wave decomposes into two circular modes, one left-handed and one right-handed (RCP and LCP, respectively) as commonly found in the study of achiral and chiral interfaces [1]. The overall phenomenon of circular polarization is also classified under the broader concept of optical activity, whereby the field vector rotations are dependent upon the nature of incident fields and also the direction of propagation [2]. It turns out that any incident plane wave traversing in an achiral/chiral (ACC) boundary essentially morphs into two (RCP/LCP) transmitted modes that diverge spatially and two (RCP/LCP) reflected modes that are co-directional. When the second medium becomes dispersive, permittivity and permeability (μ and ε) will become frequency dependent [3]; consequently, the monochromatic analysis needs substantial modification depending on the material parameters under instantaneous frequency (ω) dependence. Hence, the standard single-frequency phasor analysis will now have to be adjusted for individual (center) wavelengths, and the overall (linear) dispersion will be modeled along the lines of Fig. 2. Applying EM (tangential) boundary conditions at the point of incidence together with the chiral constitutive relation,

$$\widetilde{D} = \hat{\varepsilon}\widetilde{E} - j\widetilde{\kappa}\sqrt{\mu_0\varepsilon_0}\widetilde{H}, \qquad (1)$$

$$\widetilde{B} = j\widetilde{\kappa}\sqrt{\mu_0\varepsilon_0}\widetilde{E} + \widetilde{\mu}\widetilde{H} \qquad (2)$$

and phase matching at the boundary, one obtains the following (non-magnetic) Snell's laws for first interface under chirality and material dispersion of dielectric permittivity for the RCP and LCP modes respectively:

$$\sqrt{\varepsilon_{r_1}}\sin\theta_i = (\sqrt{\varepsilon_{r_2}(\Omega)} + \kappa)\sin\theta_{t_1}, \quad (3a)$$

$$\sqrt{\varepsilon_{r_1}}\sin\theta_i = (\sqrt{\varepsilon_{r_2}(\Omega)} - \kappa)\sin\theta_{t_2}, \quad (3b)$$

where $\sqrt{\varepsilon_{r_1}}$ and $\sqrt{\varepsilon_{r_2(\Omega)}}$ are the refractive indices for the two media with the second medium being dispersion. Also, $\theta_i$ and $\theta_{t_{1,2}}$ are the incident and transmitted angles for the two modes respectively, and $\kappa$ is the dimensionless chirality parameter. Equations (3a) and (3b) may be applied on the spherical surface at the point of the incidence both at the first and subsequently the second surface of a general thick lens configuration with radii of curvature $R_1$ and $R_2$ respectively and thickness $d$ to derive an equivalent two sets ABCD matrices for right- and left-circularly polarized modes for the entire chiral thick lens. The problem of propagation of 2D image through dispersive non chiral has been investigated in previous work [4]. Continuing the examination, it was decided to trace the image locations and magnifications corresponding to a 2-D input transparency through dispersive and chiral thick lens. The geometry of the system is shown on Fig.1. In this work, 2D imaging is examined for dual images formed via propagation through the chiral dispersive lens, assuming meridional plane and paraxial approximation, and investigating conditions for real and virtual images. In this paper, attention is focused on the tracking of image location and magnification based on dispersive permittivity and ABCD matrices [5]. An example of a 2-D polychromatic transparency is considered with the incident EM wave propagating through the dispersive thick lens. When the second medium is both dispersive and chiral, two key effects occur. First, the image planes become axially separated due to the assumed first-order dispersion. Additionally, because both transverse and axial distances depend on frequency, the images formed by different wavelengths are magnified differently. This results in variations in pixel size as well as their transverse and axial positions. Second, due to chirality, a 2D object transforms into two distinct images, one observed under (LCP) and the other under (RCP) after passing through the lens

## 2. ABCD Matrix for Chiral Dispersive Thick Lens

Referring to Fig.1, $C_1$ and $C_2$ are centers of curvature and $V_1$ and $V_2$ are vertices of the two spherical surfaces [6]. Using forward propagation across the interfaces only, the thick lens ABCD matrices for (RCP and LCP) under material dispersion derived by combination of the first and second interfaces transfer, matrices along with the propagation matrix (inside the lens) [7].

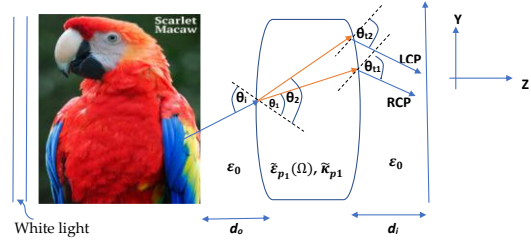

Fig.1. 2D point image incident at a chiral-chiral interface showing propagation inside the thick lens emerging to the right toward the chiral-a chiral interface.

After some steps and simplifications, the two sets of chiral dispersive thick lens ABCD matrices are given by:

$$\begin{pmatrix} A & B \\ C & D \end{pmatrix} = \begin{pmatrix} 1 - \frac{D_2 d}{n_{p2}} & \frac{d}{n_{p2}} \\ \frac{D_2 D_1 d}{n_{p2}} - D_1 - D_2 & 1 - \frac{D_1 d}{n_{p2}} \end{pmatrix} \quad (4)$$

where, the refracting power of the spherical surface $D_1$ and $D_2$ are given by:

$$D_1 = \frac{1}{R_1}(n_{p2} - n_{p1}), \quad (5)$$

$$D_2 = \frac{1}{R_2}(n_{p1} - n_{p2}). \quad (6)$$

The dispersive and chiral dispersion are manifested via the phase index $n_{p2}$ of the lens as:

$$n_{p2_{LCP}} = \left(\sqrt{\frac{(\tilde{\varepsilon}_{2_0} + \Omega \tilde{\varepsilon}'_{2_0})}{\varepsilon_0}} - \kappa_{p_0}\right) \quad (7a)$$

$$n_{p2_{RCP}} = \left(\sqrt{\frac{(\tilde{\varepsilon}_{2_0} + \Omega \tilde{\varepsilon}'_{2_0})}{\varepsilon_0}} + \kappa_{p_0}\right) \quad (7b)$$

An expression for the focal length $f$ of the chiral thick lens may be derived from the above ABCD matrix and shown to be:

$$\frac{1}{f} = \frac{(n_{p2}(\Omega,\kappa) - n_{p1})}{n_{p1}}\left[\left(\frac{1}{R_1} - \frac{1}{R_2}\right) + \frac{d}{n_{p2}(\Omega,\kappa)}\left(\frac{(n_{p2}(\Omega,\kappa) - n_{p1})}{R_1 R_2}\right)\right]. \quad (8)$$

The focal lengths, image locations, and magnifications become chiral and wavelength dependent. As a result, for any arbitrary observation screen placement behind the lens, there is no unique image plane Hence, any arbitrary plane in this system will exhibit defocusing, in axial and in transverse direction. Furthermore, the presence of chirality in the system causes two distinct images to appear on the observation screen [8].

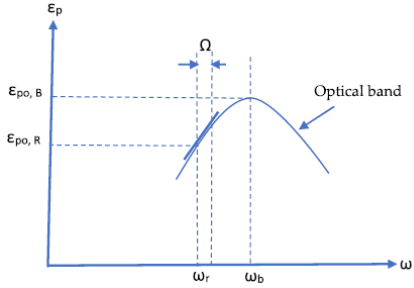

Fig. 2. Total dispersive permittivity versus instantaneous frequency [5].

## 3. Numerical Simulations, Results, and Interpretations

The derivations are utilized to analyze dual image formation for a 2D object. The plots in Fig. 3 and Fig. 4 illustrate the results. In the case of LCP, for the central red wavelength, the focal length exhibits a positive polarity within the $\tilde{\kappa}$ range of 0 to 1.8, where resonance occurs, and then transitions to negative beyond 1.8. Similarly, for the central green wavelength, the focal length remains positive between $\tilde{\kappa} = 0$ and 1.6 before switching to negative beyond 1.6. For the central blue wavelength, the focal length is positive in the $\tilde{\kappa}$ range of 0 to 1.2 and becomes negative beyond 1.2, indicating the formation of real and virtual images. In contrast, in the RCP mode, the focal length remains positive for all $\tilde{\kappa}$ values across the RGB central wavelengths.

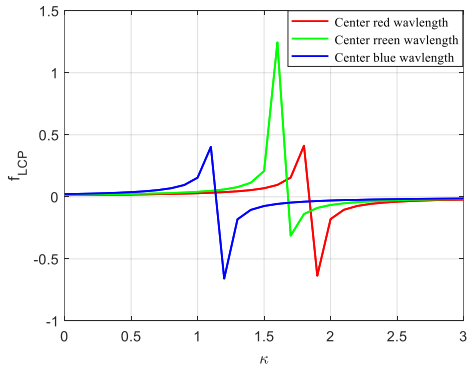

Fig.3. Focal length Vs. chirality at the center of RGB colors for LCP mode

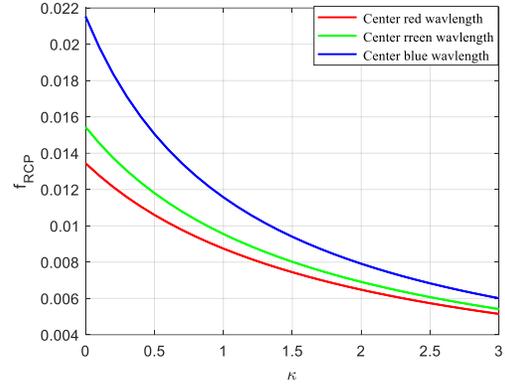

Fig.4. Focal length Vs. chirality at the center of RGB colors for RCP mode

## 3.1 Numerical Simulation of 2D image properties through Chiral-dispersive lens

We present numerical simulations of the image properties of a two-dimensional color transparency as an electromagnetic wave propagates through a lossless, *dispersive chiral thick lens.* The transmitted image is recovered on a screen positioned at an arbitrary location to the right of the lens. As shown in Fig. 5 and Fig.6, and the screen is positioned at the red image plane, both images are real. The red color components of the input image appeared with the greatest clarity and most distinctly visible across the entire image, including the background. In the case of LCP mode, the red color components are more visible than the case of LCP. When the observation screen placed in the green image planes for the RCP and LCP modes, the reconstruction images are real and the green color components of the input image is appeared with the high clarity and be most distinctly visible across the entire image in case LCP than RCP mode. When the observation screen is placed at the blue image planes, one image is real for the RCP mode and virtual for the LCP mode. The blue components are more visible in the LCP mode than in the RCP mode as seen in Fig 8 and Fig. 9.

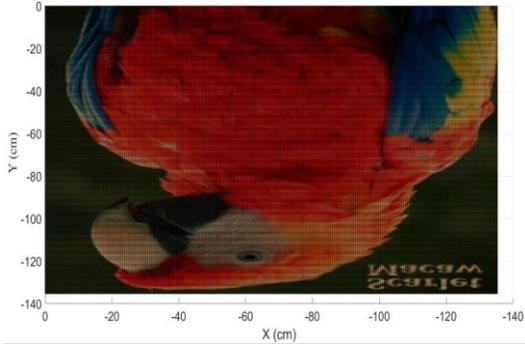

Fig.5. Reconstructed output image for RCP mode on the screen plane located at blue distance *diR*.

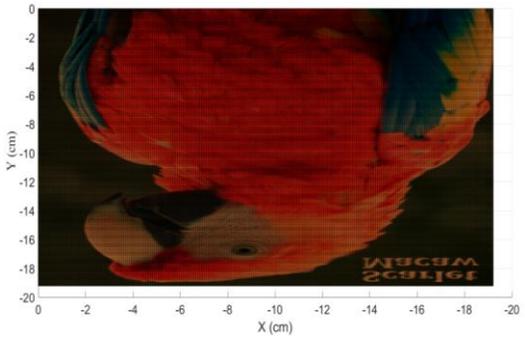

Fig.6. Reconstructed output image for LCP mode on the screen plane located at red distance *diR*.

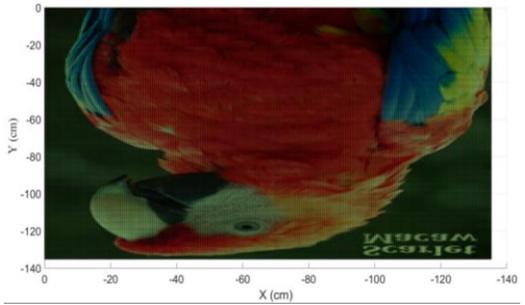

Fig.7. Reconstructed output image for RCP mode on the screen plane located at green distance *diG*.

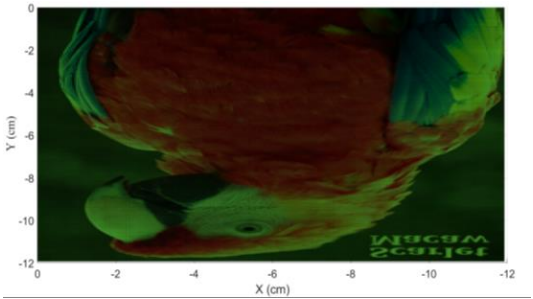

Fig.8. Reconstructed output image for LCP mode on the screen plane located at green distance *diG*.

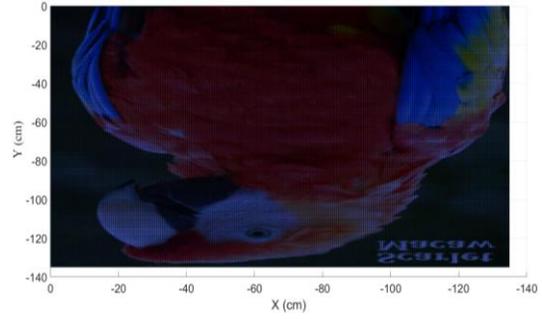

Fig.9. Reconstructed output image for RCP mode on the screen plane located at blue distance *diB*.

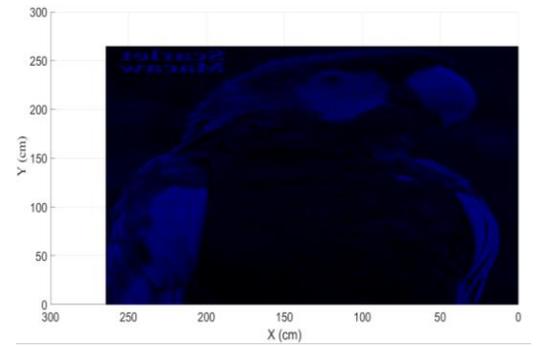

Fig.10. Reconstructed output image for LCP mode on the screen plane located at blue distance *diB*.

## 4. Conclusion

This work has explored the dual-image formation properties of two-dimensional objects propagating through a chiral-dispersive thick lens, taking into account both material dispersion and chirality effects. By developing and applying frequency-dependent ABCD matrices for right- and left-circularly polarized (RCP and LCP) modes, the study has demonstrated the axial separation of image planes and magnification variations caused by dispersive and chiral influences. Numerical simulations validated theoretical models, showing that image clarity and focus strongly depend on both the polarization state and the observation screen placement, with distinct behaviors observed across different wavelengths. Particularly, LCP and RCP modes exhibited notable differences in focal length behavior, resulting in real and virtual image formations at different chirality bands. Moreover, the color components of the reconstructed image exhibited greater clarity and were more distinctly visible across the entire image, including the background, in the case of

LCP mode compared to RCP mode. These findings provide new insights into the design and application of optical systems employing chiral materials, offering potential improvements in imaging technologies where control over polarization and dispersion is critical.